\documentstyle[11pt,newpasp,twoside,epsf]{article}
\def\edcomment#1{\iffalse\marginpar{\raggedright\sl#1\/}\else\relax\fi}
\reversemarginpar

\begin{document}
\title{Globular Cluster Distances and Physical Properties from 
Double- and Fundamental Mode Variables}
 \author{G\'eza Kov\'acs}
\affil{Konkoly Observatory, P.O. Box 67, H-1525, Hungary}

\begin{abstract}
We briefly describe a method of obtaining relative intrinsic stellar 
parameters, such as absolute magnitudes and colors from the light 
curves of fundamental mode RR~Lyrae (RRab) stars. Absolute stellar 
parameters of double-mode variables (Cepheids and RR~Lyrae stars) 
can be computed from their periods and intrinsic colors combined 
with stellar atmosphere and pulsation models. By using these two 
methods, one can determine, e.g., the distance of any RRab star, 
assuming that accurate multicolor photometry is available. By applying 
this and a related method for computing metallicity, we derive 
distance, reddening and metallicity distribution for the RRab stars 
of $\omega$~Cen. 
     
\end{abstract}


\section{Introduction}

RR~Lyrae stars have played a very important role in astrophysics since 
the end of the nineteenth century, when the first photographic plates 
on the variables of globular cluster M5 were taken by Bailey (1899). 
One of their best known (and perhaps most frequently utilized) properties 
is the relatively small spread in their average luminosities. The range 
of this quantity is about $0.4$~mag among cluster variables and remains 
under $0.6$~mag even if we include more metal rich RR~Lyrae stars, such 
as the ones in the Galactic field. In the last decade or so many studies 
have been made to measure/compute RR~Lyrae luminosities more accurately. 
Although it is true that metallicity plays a significant role in determining 
the luminosity, evolutionary effects should also be considered together 
with other unknown parameters, most often referred to as `second parameters'. 

Several years ago we started a program aimed at deriving {\it relative} 
intrinsic parameters of fundamental mode (RRab) variables from their light 
curves. This idea has been pursued at various levels from the very early 
dates of pulsating star studies. The best known among them are the various 
period--luminosity and period--luminosity--color relations. In a more 
advanced approach, amplitudes have also been involved (e.g., Sandage 1981). 
We extended these studies to include quantities representing the shape of 
the light curve. It turned out that some of these quantities (i.e., 
certain basic Fourier parameters) significantly improve the regressions 
to some intrinsic parameters, such as the absolute magnitude $M_V$ and 
iron abundance [Fe/H]. Our approach is completely {\it empirical} and 
yields accurate estimates for the relative stellar parameters. In Sect.~2 
we give a brief description of the method and summarize the updated 
formulae. 

Considering absolute distances of various stellar systems, we studied 
double-mode variables which have the distinct advantage of yielding 
extra constraints on the stellar parameters through their two observed 
pulsation modes which are identified with two of the first three radial 
eigenmodes. By using the very recent data sets from the {\sc macho} and 
{\sc ogle} variable star inventories, with the aid of pulsation and 
stellar atmosphere models we derived the distances of several globular 
clusters and those of the Magellanic Clouds. Sect.~3 gives a concise 
description of the method and a summary of the derived distances. 

Unfortunately, direct application of the above method to $\omega$~Cen is 
not possible, because of the mysterious lack of double-mode variables in 
this, otherwise well-populated cluster. Therefore, in Sect.~4 we use 
relative cluster distances to derive the absolute distance of $\omega$~Cen. 
We also estimate its overall reddening by using two independent data sets. 
A comparison is made between our distance and the one derived from 
the analysis of the detached binary OGLEGC17 (Thompson et al. 2000; 
see also Kaluzny 2002). It is shown that there is a significant difference 
of $0.2$~--~$0.4$~mag between the two distances, with our method favoring 
the `long distance scale'. 

The often quoted peculiar metallicity distribution of this cluster is 
investigated in Sect.~5. From a representative sample of RRab stars, 
our derived metallicity distribution shows a fair agreement with the 
ones obtained from the spectroscopic investigation of the giant 
population. Furthermore, it is pointed out that there are other 
clusters too, where we may suspect sufficient metallicity spread 
based on the light curves of the RRab stars.


\section{Relative intrinsic quantities and metallicities from RRab stars}

The underlying assumption in using the light curves of pulsating stars 
in calculating stellar parameters is that there exist a `reasonably simple' 
relation between the Fourier and stellar parameters. Unfortunately, except 
for a broad understanding of this dependence from hydrodynamical models, 
there is no precise theoretical prediction on the functional form of 
this relation. Therefore, we approached the problem by searching for 
best fitting {\it linear} combination of the Fourier parameters. All of 
our results suggest that although the existence of nonlinear relations 
cannot be excluded, they might play a role only at a level which is 
lower than the estimated observational noise. 

Since our method has been described in several papers (e.g., Kov\'acs \& 
Jurcsik 1996; 1997; Kov\'acs \& Walker 2001, hereafter KW01), here we 
mention only the most important steps. In the context of globular cluster 
RRab stars, we consider the relation between the observed average 
magnitudes and the Fourier parameters of the light curves in the Johnson 
$V$ band. Because we deal with different clusters, it is also necessary 
to find the best relative ({\it reddened}) distance moduli. This is done 
within the fitting procedure, where the Fourier parameters and the 
relative distance moduli are fitted simultaneously to the apparent 
average magnitudes. It was shown in KW01 on the basis of 366 stars 
in 20 clusters that the optimum fit to the {\it intensity-averaged} 
magnitudes is obtained by the following three-parameter formula 
%
%
\begin{eqnarray}
M_V = -1.876\log P - 1.158A_1 + 0.821A_3 + const.
\end{eqnarray}
Here $M_V$ denotes the absolute visual magnitude, $P$ is the period 
(in [day]), $A_1$ and $A_3$ are the Fourier amplitudes (in [mag]) of 
the fundamental and second harmonic components, respectively. This  
formula (with the corresponding relative distance moduli) fits the 
above mentioned data set with a standard deviation of $0.040$~mag, 
which is higher than the expected observational noise. As it is 
discussed in KW01, some part of the dispersion can be accounted for 
by the reddening variation in front of the clusters. Nevertheless, 
even if we consider this effect, the estimated overall observational 
noise of the individual average magnitudes becomes $\approx 0.02$~mag, 
which is perhaps still higher than expected. We will have a better 
understanding of this problem when future precise {\sc ccd} observations 
will become available on clusters with low reddenings. 

We can suppress the effect of reddening variation by using two-color 
data, and construct a {\it reddening-free, magnitude-averaged} quantity, 
such as $W=V-3.1(B-V)$ (see Dickens \& Saunders 1965). By following the 
same procedure as for $M_V$, with the aid of 172 stars in 15 clusters, 
KW01 obtained the following formula       
%
%
\begin{eqnarray}
V_0 - 3.1(B-V)_0 = -2.467\log P + const.
\end{eqnarray}
We see that for this particular quantity, beside the period, there seems 
to be no need for any other information on the light curve. As it is 
described in KW01, this result is in contradiction with our earlier 
finding, which, however was based on the analysis of a much smaller 
data set. In addition, when a comparison is made with the regression 
on the reddening-free color index $Q=V-I-1.24(B-V)$ computed from a 
sample including also Galactic field variables, it turns out that 
either $V-3.1(B-V)$ or $V-2.5(V-I)$ should contain some dependence 
on $\varphi_{31}$ (as it is expected from the metallicity dependence of 
$Q$ and from the $\varphi_{31}$ dependence of [Fe/H] -- see Jurcsik \& 
Kov\'acs 1996). None of these quantities, based on cluster variables 
show this dependence. We explain this contradiction as a result of the 
relatively high noise and small metallicity range of the available 
cluster variables. 

Eq.~(2) (with the computed {\it true}, i.e., {\it reddening-free} 
relative distance moduli) fits the data with the standard deviation 
of $0.041$~mag. We see that this is almost the same value as the one 
obtained in the fit of the average $V$ magnitude, although we would 
expect a smaller one, because individual reddenings are thought to 
be eliminated in the color combination used. However, there is a 
`price to pay' for this, namely the observational noise is amplified, 
which, in the case of cluster variables, yields a rather large 
contribution to the overall error budget.  

With the combination of the {\it magnitude-averaged} counterpart of 
Eqs.~(1) (see KW01) and Eq.~(2), with the zero point given by 
Kov\'acs \& Jurcsik (1997) we obtain the following expression for 
the intrinsic color index
%
%
\begin{eqnarray}
(B-V)_0 = 0.189\log P - 0.313A_1 + 0.293A_3 + 0.460
\end{eqnarray}
With this expression one can easily estimate the reddening without 
resorting to some more elaborate formulae (e.g., Blanco 1992), which 
are based on less extended samples and use additional quantities, 
such as metallicity. If less accurate data are available, one can 
also use a two-parameter formula with $P$ and $A_1$ as given in KW01 
(of course, in this case the result will be somewhat less accurate). 

Finally we quote our earlier result concerning the dependence of 
[Fe/H] on the Fourier parameters. Based on a sample of 81 Galactic 
RRab stars, Jurcsik \& Kov\'acs (1996) found that the published 
iron abundances, obtained from low-dispersion spectroscopic data, 
can be fitted within the expected accuracy of the measurements with 
the following formula
%
%
\begin{eqnarray}
[{\rm Fe/H}] = -5.394P + 1.345\varphi_{31} - 5.038
\end{eqnarray}
The standard deviation of the fit is 0.13~dex, and the contribution 
from the Fourier phase is highly significant. The data span almost 
the full expected range of RR~Lyrae metallicities, i.e., from 
[Fe/H]=$-2.0$ to $0.0$. The fact that the metallicity of RR~Lyrae 
stars depends also on other parameters, beside the period, was first 
recognized by Preston (1959). Although the period-amplitude diagram 
is still used for diagnostic purposes by some researchers, it is 
important to note that Eq.~(4) fits the above mentioned data set 
with a factor of {\it two} lower dispersion than the regression 
employing $P$ and $A$. Application of Eq.~(4) to globular clusters 
yielded good agreement with the spectroscopic values obtained from 
giants. At the low-metallicity end, however, our formula predict 
a somewhat (by $\approx 0.2$~dex) higher metallicity. Although it 
is possible that Eq.~(4) overestimates the metallicity in the low 
[Fe/H] region, because of the relatively high number of less metal 
poor variables in the Galactic field, it is important to note that 
a better comparison would require direct spectroscopic measurements 
for cluster RR~Lyrae stars. However, these data, with a comparable 
accuracy to that of the Galactic field stars, practically do not exist.   


\section{Absolute stellar parameters from double-mode variables}

Due to the current microlensing surveys in the Magellanic Clouds, 
the number of known double-mode stars (RR~Lyrae and $\delta$~Cephei 
stars pulsating in two radial modes simultaneously) has increased 
considerably. Based on the analysis of 1350 variables, previously 
classified as first overtone RR~Lyrae (RRc) stars, Alcock et al. 
(1997, 2000) discovered 181 double-mode (RRd) stars in the Large 
Magellanic Cloud (LMC). From the `fall-out' of the {\sc ogle} project, 
now we know 76 double-mode Cepheids in the LMC and 93 ones in the Small 
Magellanic Cloud (SMC) (Udalski et al. 1999a, Soszy\'nski et al. 2000). 
While in the case of RRd stars the so far securely identified variables 
all show pulsations only in the fundamental and first overtone modes, 
we find variables in great number among Cepheids both in the LMC and SMC 
which pulsate in the first and second overtones. For the significance 
of these discoveries, it is enough to mention that prior to the 
microlensing surveys, it was CO~Aur known as the only example for first 
and second overtone pulsation among Cepheids (Babel \& Burki 1987). 
Now, according to the {\sc ogle} surveys mentioned above, the number 
of these variables in the Magellanic Clouds is 127. 

The importance of double-mode variables comes from fact that their 
observed periods are very close to the ones computed from linear, 
purely radiative pulsation models (i.e., the change in the periods 
due to nonlinear and convective effects is very small -- see Koll\'ath 
\& Buchler 2001). Therefore, the periods are easily comparable with 
large number of models, and the stellar parameters can be strongly 
constrained. Since the radial normal mode periods depend mainly on 
{\it four} parameters (mass, luminosity $L$, effective temperature 
$T_{\rm eff}$, and heavy element abundance [M/H]), the two periods yield 
the following expression for the luminosity 
%
%
\begin{eqnarray}
\log L = f(P_0,P_1,\log T_{\rm eff},{\rm [M/H]})
\end{eqnarray}
Here $f$ is a smoothly varying function, usually not computed 
explicitly, but obtained from pulsation model grids through some 
interpolation method. Mass is derived in a similar manner. 

For the computation of $T_{\rm eff}$ we can use the observed dereddened 
color index and convert it to $T_{\rm eff}$ with the aid of some current 
calibration using infrared flux method or other, least 
model-dependent methods (see, e.g., Blackwell \& Lynas-Gray 1994; 
Alonso, Arribas \& Mart\'\i nez-Roger 1999; Sekiguchi \& Fukugita 2000). 
In our approach we used these more empirical methods to fix the 
{\it zero point} of the temperature scale, but the actual functional 
dependence of $T_{\rm eff}$ on color index, $\log g$ and [M/H] is obtained 
from grids of model atmospheres (Castelli, Gratton \& Kurucz 1997). 
Metallicity [M/H] is usually identified with the iron abundance [Fe/H] 
and obtained either from independent estimates or from the RRab population 
of the host cluster as described in Sect.~2. 

Various types of variables have been used to test the above method in 
deriving distances to nearby globular clusters and to the LMC. In the 
case of RRd stars we used $B-V$, whereas for Cepheids we employed $V-I$ 
color indices, because they were the ones which were most numerous in 
the corresponding publications. For the $T_{\rm eff}$ zero points we used 
the calibration of Blackwell \& Lynas-Gray (1994). A large number of 
purely radiative linear pulsation models were computed in the standard 
way. We used the {\sc opal} opacities as given by Iglesias \& Rogers 
(1996). Additional details of the databases and model fitting can be 
found in Kov\'acs (2000a, b) and references therein. 
%
%
%
\begin{table}
\caption{Consistency of the true distance moduli obtained from 
double-mode variables}
\begin{center}
\begin{tabular}{||lrrc||}
\hline 
Cluster & $N$ & [Fe/H] & $DM_{LMC}$ \cr
\hline\hline
\phantom{$^{\star}$}M~15     &   8 & $-2.3$ & 18.52     \ \ \ \ \ \    \\
\phantom{$^{\star}$}M~68     &  11 & $-2.0$ & 18.47     \ \ \ \ \ \    \\
\phantom{$^{\star}$}IC~4499  &  13 & $-1.5$ & 18.50     \ \ \ \ \ \    \\
\phantom{$^{\star}$}LMC      & 181 & $-1.5$ & 18.52     \ \ \ \ \ \    \\
         $^{\star}$SMC       &  93 & $-0.8$ & 18.54     \ \ \ \ \ \    \\
         $^{\star}$LMC       &  71 & $-0.3$ & $^+$18.58 \ \ \ \ \ \ \ \ \\
\hline
\end{tabular}\\ [7pt]
\parbox{67mm}{
\footnotesize {$^{\star}$Beat Cepheid distances} \\
\footnotesize {$^+$Preliminary result}
}
\end{center}
\end{table}
By using relative distances derived from the RRab populations as described 
in Sect.~2 and $DM_{LMC}-DM_{SMC} = -0.51$ as given by Udalski et al. 
(1999b), we converted the cluster distances to the distance modulus of the 
LMC. The result is shown in Table~1. The the near agreement of the 
distances derived from diverse stellar populations is very comforting and 
shows that: 
(i) the photometric data used from various sources have consistent 
zero points; 
(ii) relative distances are well-calibrated; 
(iii) overall metal abundances (which are at the same time in agreement 
with the generally accepted cluster or population averages) are 
properly selected --- double-mode masses and luminosities are sensitive 
functions of the metal abundance; 
(iv) models yield consistent results for very different double-mode 
pulsators (RR~Lyrae, first overtone/fundamental and first 
overtone/second overtone Cepheids). 

Although the method is affected by various errors, most of 
them cause changes in the derived distances at a level of several 
hundredths of a magnitude. Statistical errors are especially small, 
because of the large number of objects available for us from the 
microlensing surveys and also because of the notable accuracy 
of the globular cluster data. Certain systematic errors (e.g., the 
ones due to photometry, and metal abundance) must also yield a rather 
small contribution to the total error budget, because otherwise we 
would get larger differences among the distances derived from 
various stellar populations. However, there is one source of error 
which affects all distances in almost the same way, and thereby should 
be considered as the major ambiguity of the method. This is the choice 
of the zero point of the color index~$\rightarrow T_{\rm eff}$ calibration. 
In our studies we took the temperature scale based on the infrared 
flux method as given by Blackwell \& Lynas-Gray (1994). We chose their 
zero point because we found their result and the adjacent data published 
by Clementini et al. (1995) to be most easily used in fixing the zero 
points of the relative $T_{\rm eff}$ values of the atmosphere models of 
Castelli et al. (1997). By using the zero points of other recent 
calibrations (e.g., Alonso et al. 1999; Sekiguchi \& Fukugita 2000) 
the distances shown in Table~1 change by amounts less than $-0.10$~mag. 
This puts a {\it lower} limit of $\approx 18.4$~mag for the true distance 
modulus of the LMC. This is significantly different from the distance 
derived from, e.g., binary stars (Guinan et al. 1998) and red clump stars 
(Udalski et al. 1999b), but in agreement with the distance obtained 
from the Baade-Wesselink analysis of Cepheids (Gieren, Fouqu\'e \& 
G\'omez 1998) and evolution models of Cepheids and RR~Lyrae stars 
(Alibert et al. 1999; De~Santis \& Cassisi 1999; VandenBerg et al. 
2000). We will return to the distance problem in the next section.  

The average physical parameters of the RRd stars obtained from the 
above analyses show that the main difference between the variables 
hosted by low- and high-metallicity clusters is that the high-metallicity 
stars have slightly higher temperatures than the low-metallicity ones. 
Differences between the luminosities and masses are insignificant. For 
example, for IC~4499 and M~68 we get 
$0.755\pm0.021$, $1.708\pm0.017$, $3.830\pm0.004$ and 
$0.756\pm0.016$, $1.728\pm0.013$, $3.820\pm0.004$, for the average 
$M/M_{\sun}$, $\log L/L_{\sun}$ and $\log T_{\rm eff}$ values, respectively. 
The errors listed are the standard deviations calculated from the 
values of the individual variables (i.e., they are not the errors of 
the averages, those are about a factor of three smaller). It follows 
from this result that the observed trend/spread in the 
$P_0\rightarrow P_1/P_0$ diagram in the LMC, Sculptor and Sagittarius 
dwarf galaxies (Alcock et al. 2000; Kov\'acs 2001; Cseresnjes 2001), 
is most probably due to a spread in the metallicity rather than in the 
mass of the RRd populations of these systems. Although the above 
stellar parameters were derived by using solar-type element ratios for 
the heavy elements, we think that the relative parameters will remain 
unchanged even if, e.g., oxygen enhancement is included. This is because 
in the metallicity range of $(-2.2, -1.3)$, relevant for RRd stars, 
the enhancement ratio does not change significantly (e.g., Clementini 
et al. 1995). Unfortunately, without the availability of detailed 
abundance analyses on RRd stars, we cannot make a more quantitative 
statement on the affect of the various element enhancements. 
Nevertheless, the agreement with the distance moduli obtained from 
Cepheids, which have scaled solar heavy element mixtures, implies 
that enhancement of some components (other than iron) in the total heavy 
element content $Z$ do {\it not} play a significant role in the finally 
derived physical parameters.


\section{The distance and reddening of $\omega$~Cen}

First we compute the relative distance of $\omega$~Cen. 
In order to do so, we need multicolor photometry which is extended 
and accurate enough to allow us a reliable computation of the average 
colors and Fourier parameters. Unfortunately, among the publicly available 
data sets, the quite recent {\sc ccd} observations of Kaluzny et al. (1997) 
(hereafter K97) were made only in $V$ color. Therefore, it was necessary 
to employ other, less contemporary data sets. Nevertheless, we will see 
that these older data yield consistent and accurate estimates on the 
average distances and reddenings. 

The two data sets used are those of Dickens \& Saunders (1965) 
(hereafter DS65) and Butler, Dickens \& Epps (1978) (hereafter B78). 
Both of these papers contain data on the {\it intensity mean} averages 
(i.e., magnitude averages computed from the mean of the intensity 
transformed light curves). Because the distance modulus and reddening 
formulae (Eqs. (2), (3)) refer to {\it magnitude mean} averages 
(i.e., averages of the standard magnitude values), it is necessary 
first to convert the published intensity averages $\langle V\rangle$, 
$\langle B\rangle$ to magnitude averages $\overline V$, $\overline B$. 
This is done for variables common with the sample of K97  
by using the Fourier parameters of the accurate $V$ light curves of K97. 

To derive the relations between the two types of averages, we used the 
$V$ Fourier parameters and the corresponding $V$ and $B$ averages of 
150 stars compiled by Kov\'acs \& Jurcsik (1997) and KW01. By 
experimenting with various number of parameters and polynomial dependence, 
we obtained the following result. Higher parameter linear combinations 
do not improve significantly the quality of the single parameter fit 
obtained with the first Fourier amplitude $A_1$
%
%
\begin{eqnarray}
\overline{V}-\langle V\rangle = -0.039 + 0.238A_1 , 
\hspace*{35mm}\sigma=0.0037
\end{eqnarray}
where $\sigma$ means the standard deviation of the fit in [mag].   
Fig.~1 shows that the correct relation must include significant 
nonlinear term(s). Indeed, the best single parameter quadratic 
fit takes care of most of the nonlinearity and yields a considerably 
lower standard deviation
%
%
\begin{eqnarray}
\overline{V}-\langle V\rangle = 0.002 - 0.040A_1 + 0.437A_1^2 , 
\hspace*{20mm}\sigma=0.0021
\end{eqnarray}
Finally, when the number of parameters is increased to two in a 
quadratic fit, containing all parameter combinations up to second 
order, we end up with the following expression
%
%
\begin{eqnarray}
\overline{V}-\langle V\rangle = 0.002-0.002A_1-0.078A_3 
\hspace*{23mm}\phantom{\sigma=0.0011}\nonumber \\
\phantom{\overline{V}-\langle V\rangle =}
+0.198A_1^2 + 0.695A_1A_3 - 0.131A_3^2, 
\hspace*{9.3mm}\sigma=0.0011\hspace*{3.3mm}
\end{eqnarray}
where $A_3$ denotes the Fourier amplitude of the second harmonics 
(the $2\omega$ component). As it is shown in Fig.~1, indeed, the above 
regression yields an excellent fit to the data, leaving no trace of 
systematic effects, except maybe for a slight increase of dispersion 
in the high amplitude regime, where the differences between the two 
types of means become larger. Even this small dispersion can be further 
decreased by employing a three-parameter quadratic fit. Although tests 
made on much larger data sets show that this high-parameter fit yields 
statistically significant improvement compared to the two-parameter one, 
for the practical purpose in the present context the two-parameter 
quadratic formula is quite satisfactory. 
%
%
%
\begin{figure}
\plottwo{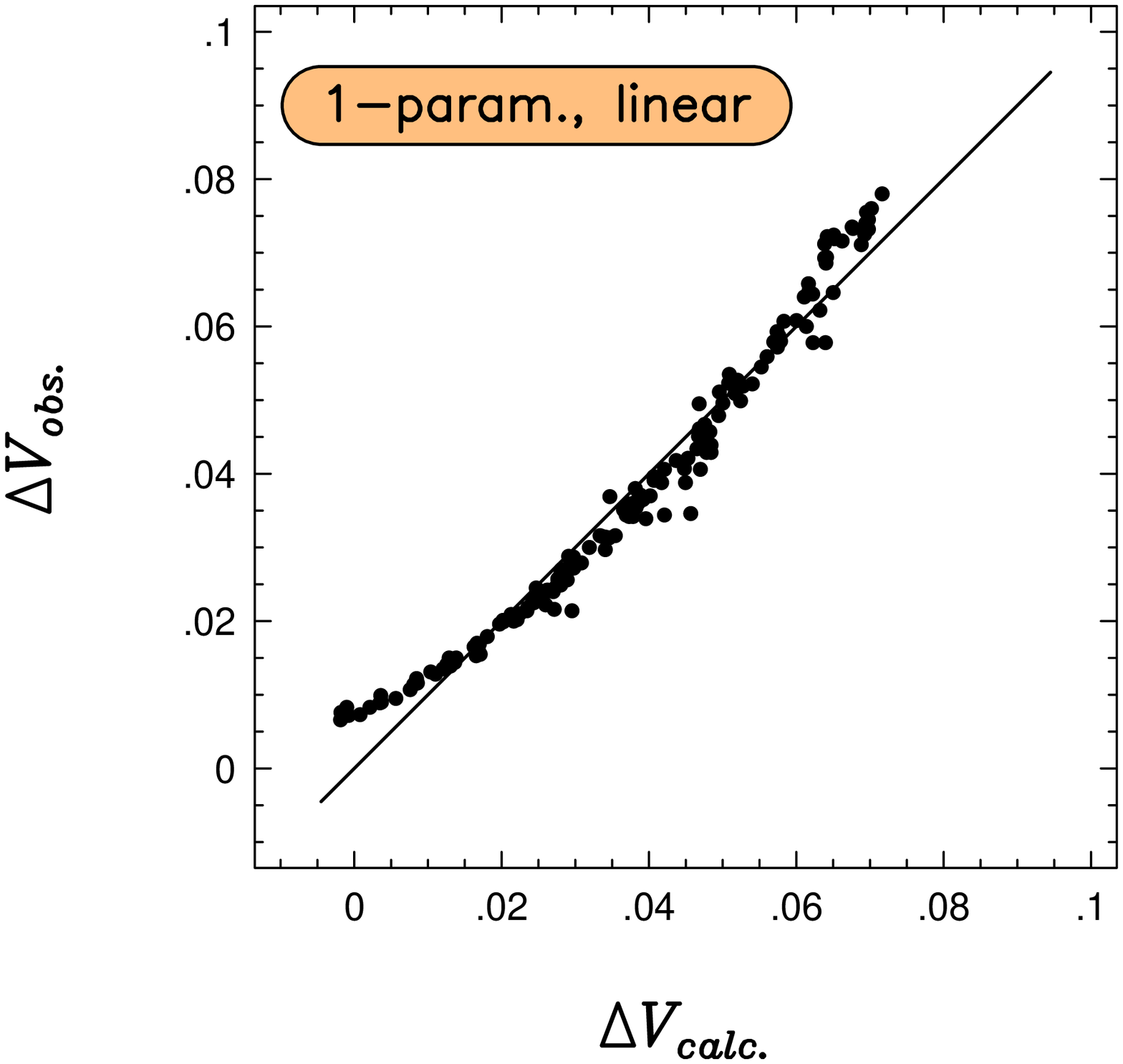}{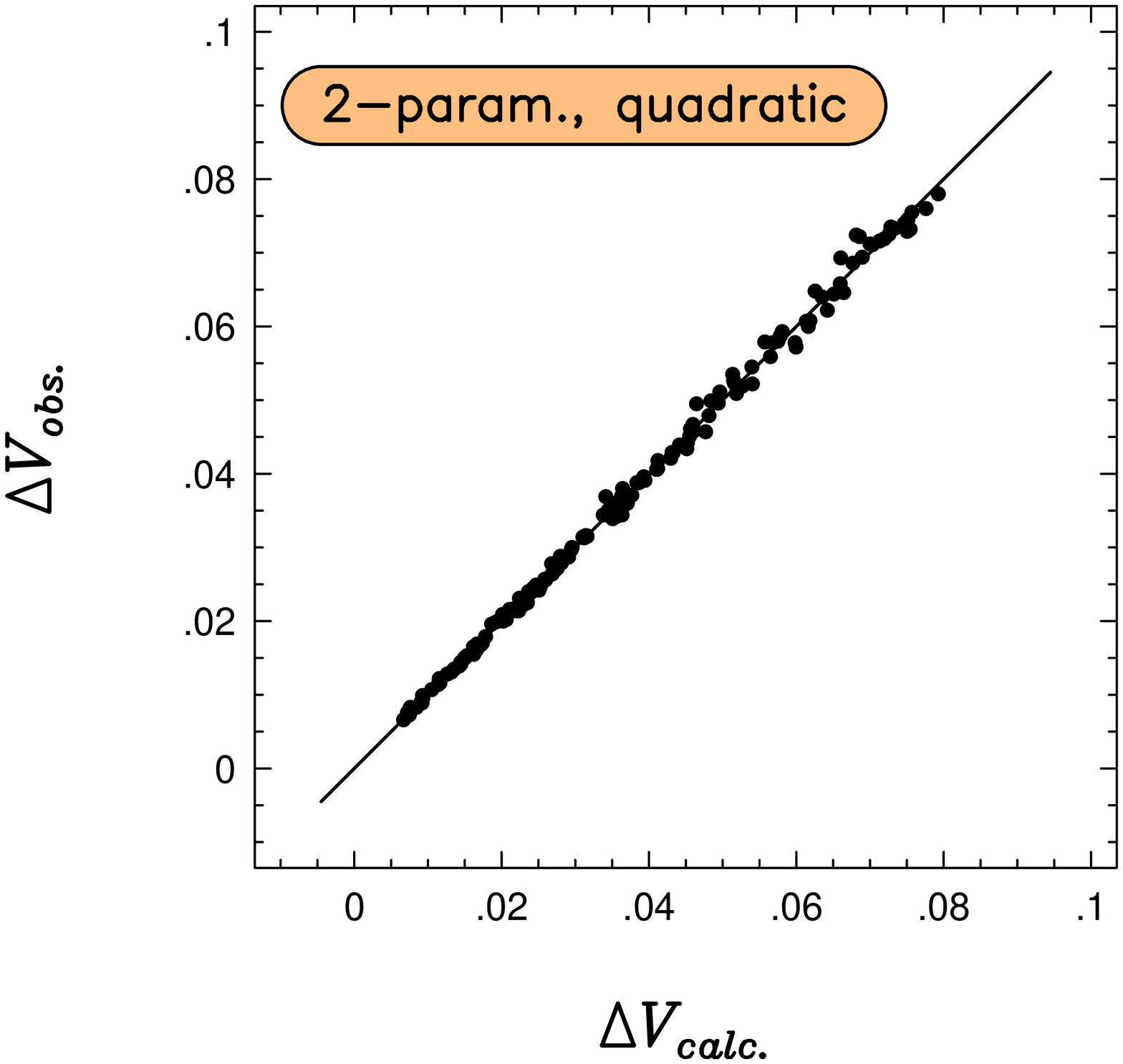}
\caption{Comparison of two regressions of the observed difference 
($\Delta V_{obs}=\overline V-\langle V \rangle$) between the magnitude 
and intensity mean averages. For the specific forms of the regressions 
we refer to Eqs. (6) and (8).}
\end{figure}

In the next step we repeat the above procedure for the $B$ averages 
defined on the same data set. We get the following formulae
%
%
%
\begin{eqnarray}
\overline{B}-\langle B\rangle 
& = & -0.067 + 0.418A_1 , 
\hspace*{32mm}\sigma=0.0081 \\
& = & \phantom{-}0.003 - 0.058A_1 + 0.752A_1^2 , 
\hspace*{14mm}\sigma=0.0061 \\
& = & \phantom{-}0.012 - 0.134A_1 + 0.056A_2 \nonumber \\
& \phantom{=} & +0.475A_1^2 + 0.794A_1A_2 - 0.175A_2^2, 
\hspace*{5mm}\sigma=0.0045
\end{eqnarray}
It is noted that nonlinearity plays a significant role also in this 
color. We see that the dispersions are higher than for $V$. This is 
because we use the Fourier parameters of the $V$ light curves also in 
this case. Nevertheless, even this larger dispersion corresponds to 
a rather tight regression if we compare it with the total range of 
$\overline{B}-\langle B\rangle$, which is $0.13$~mag. The fit becomes 
much better if the Fourier decompositions of the $B$ light curves are 
used. However, our purpose is to use the available accurate $V$ light 
curves to compute the necessary corrections. Observational errors 
are much larger in the data sets we use in this paper than 
the errors introduced by the lower accuracy of the fit of the $B$ data.  
%
%
%
\begin{figure}
\plottwo{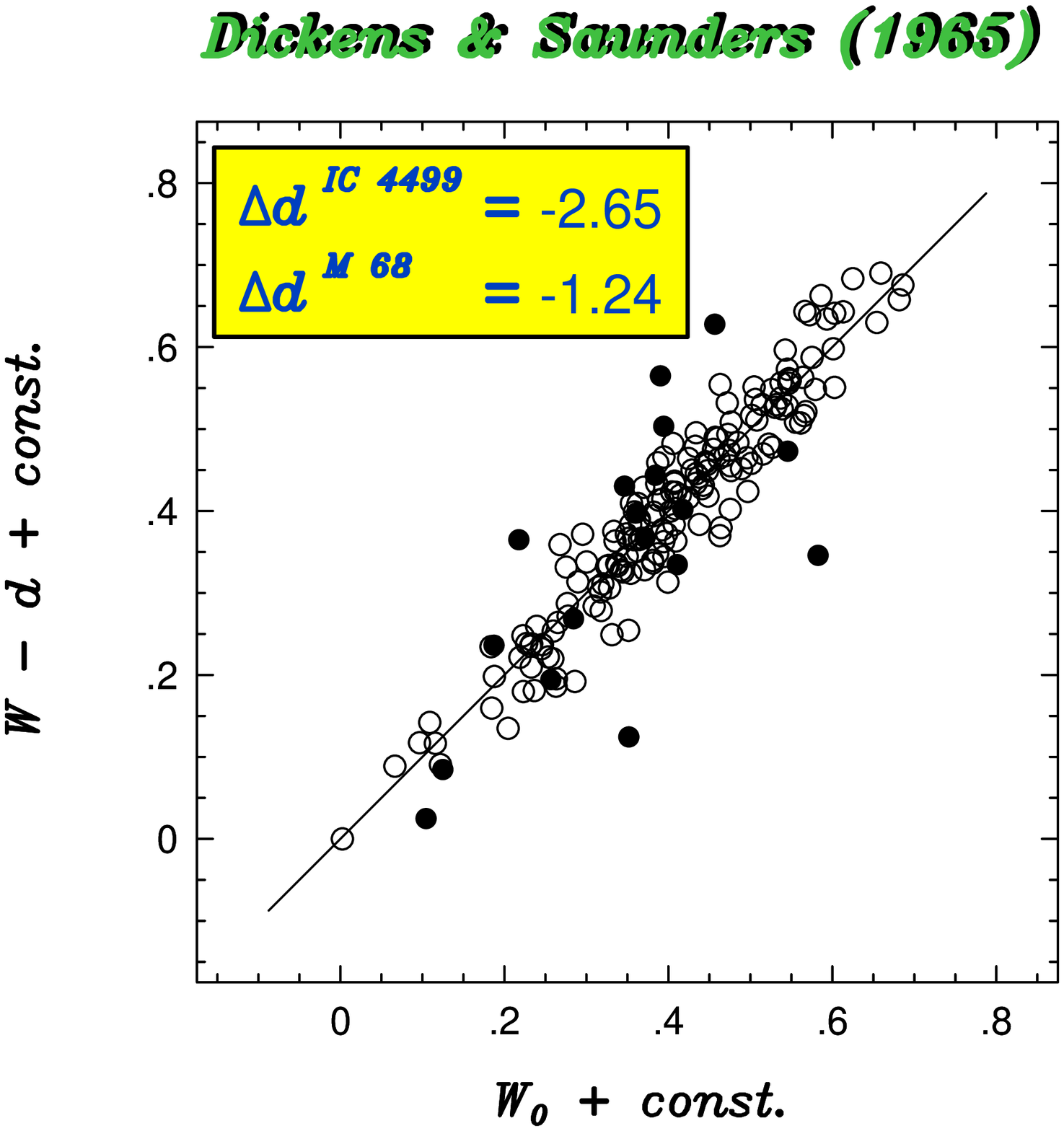}{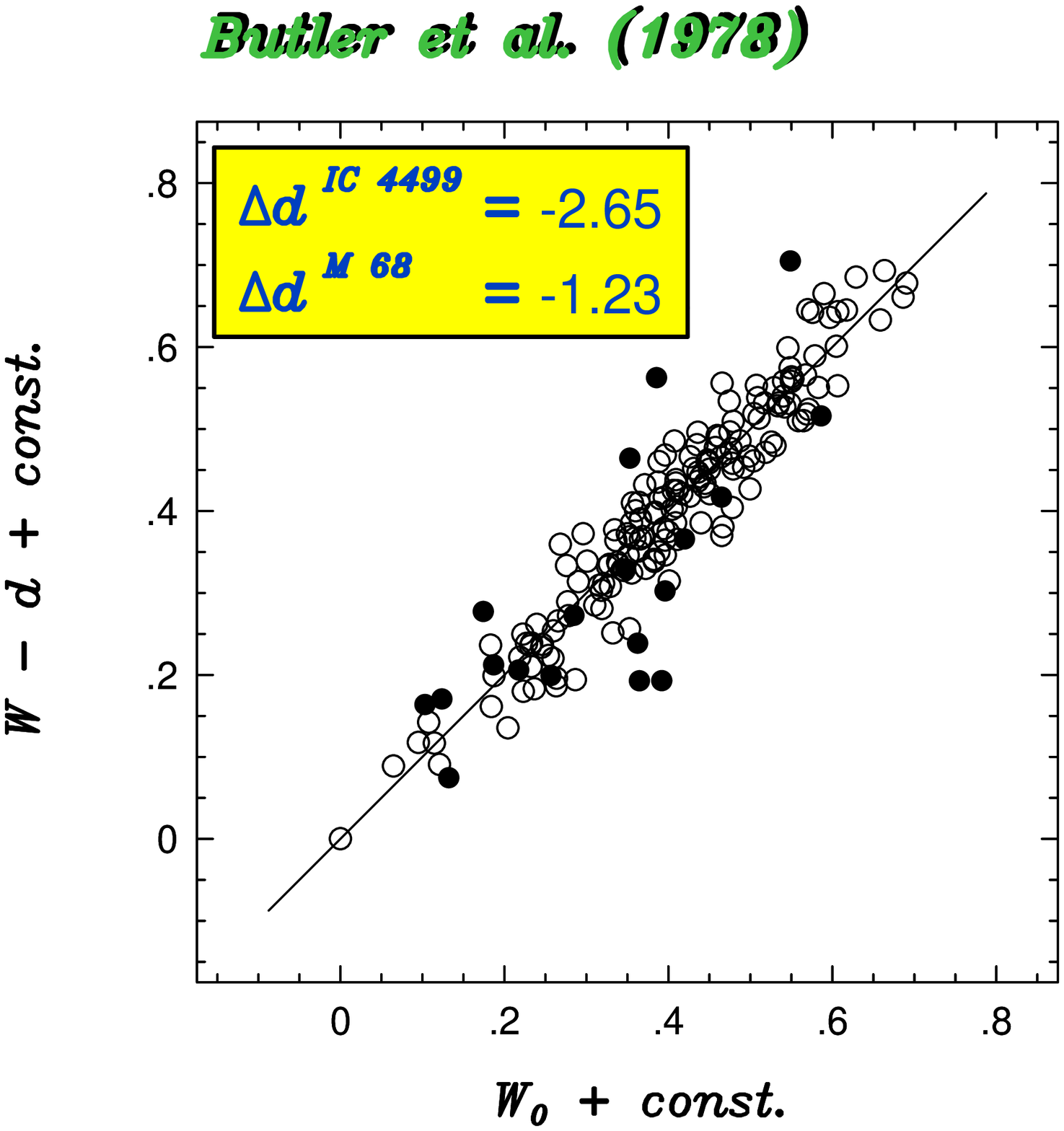}
\caption{Fit of $W=V-3.1(B-V)$ of the $\omega$~Cen RRab stars 
(filled circles) for the data sets shown in the header. Variables 
from the sample of Kov\'acs \& Walker (2001) are shown by open 
circles. Derived relative distance moduli ($\omega$~Cen {\it minus} 
cluster) are displayed in the upper left corners. The $45\deg$ lines 
are shown for reference.}
\end{figure}

By applying Eqs.~(8) and (11), we transformed the intensity mean values 
of those stars in the DS65 and B87 samples which have also been observed 
by K97. Then, Eq.~(2) can be used directly on these data to compute 
relative distance moduli. For better illustration, in Fig.~2 we plot the 
result of the direct fit of the $\omega$~Cen data to the corresponding 
data set of KW01. The few data points on M4 have been omitted, in order 
to be consistent with a subsequent fit of $V$. (The cluster M4 has a rather 
large and inhomogeneous reddening, therefore, most of its stars show up as  
outlying points in the fit of $V$.) We see that both data sets yield the same 
relative distance moduli, in spite of the large scatter of these unedited 
samples. With the same method (or by using the corresponding formula of 
KW01 for the magnitude mean $V_0$), we can get the following {\it reddened} 
relative distance moduli: $\tilde{d}_V(\omega$~Cen$)-\tilde{d}_V$~(IC~4499)$=-3.05$ 
and $\tilde{d}_V(\omega$~Cen$)-\tilde{d}_V$~(M~68)$=-1.06$ for the 
18 variables of DS65 and $-3.04$, $-1.05$ for the 20 variables of B78. 
We note that in KW01 we obtained reddened, relative distance moduli of 
$-2.99$ and $-1.01$. The consistency of the DS65 and B78 distance 
moduli indicates that there is a zero point error of about 
$0.03$--$0.06$~mag in the K97 data (meaning fainter magnitudes for the 
variables of K97).   

In calculating the absolute distance modulus, we simply add to the 
above relative distance moduli the absolute distance moduli of IC~4499 
and M~68, derived from their RRd populations. The distances of these 
two clusters can be computed according to Kov\'acs (2000b) (see also 
Kov\'acs \& Walker 1999). We get $16.47$ and $15.07$~mag for the true 
distance moduli of IC~4499 and M~68, respectively. Finally, with the 
above derived relative distance moduli we get 13.83~mag for the true 
distance modulus of $\omega$~Cen. 

Reddening can be computed by simply applying Eq.~(3) to the Fourier 
decompositions of the light curves of K97 and compare the calculated 
intrinsic color indices with the observed ones (after applying the 
above transformation to get magnitude averages). For the above sample 
of DS65 we get $E_{B-V}=0.092\pm 0.029$, wheras for that of B78, we 
obtain $E_{B-V}=0.095\pm 0.029$. Here the errors denote the standard 
deviations of the reddening values of the individual variables. The 
statistical error of the averages is $0.007$~mag. Although in 
Kov\'acs \& Jurcsik (1997) we used a reasonably large number of stars 
from the Galactic field and globular clusters to fix the zero 
point of the reddening scale, there might be also a systematic 
error in the above derived values, but this should not exceed 
$\pm$(0.01--0.02)~mag.

It is worthwhile to compare this distance with the one derived by 
Thompson et al. (2001) (hereafter T01, see also Kaluzny et al. (2002)). 
Their distance estimate is based on the analysis of the detached 
binary OGLEGC17. This method is similar to ours in the sense that 
in the calculation of the luminosity it is essential to adopt 
some color index$\rightarrow T_{\rm eff}$ calibration. For this reason we 
employ the $T_{\rm eff}$ scale of Sekiguchi \& Fukugita (2000), which 
is the same as the one used by T01. Furthermore, as in all of our 
previous computations, we use the more customary value of 
$M^{bol}_{\sun}=4.75$ instead of $4.70$, used by T01. By considering 
only the primary component, with $E_{B-V}=0.13$, $(B-V)=0.66$, 
$V=17.46$, [Fe/H]=$-1.74$, $R=1.882$~$R_{\sun}$, $B.C.=-0.22$, 
$T_{\rm eff}$(Sun)$=5780$~K, we get 13.38~mag for the true (dereddened) 
distance modulus $DM$. By repeating the above computation with 
$E_{B-V}=0.09$, we get almost the same result of $DM=13.41$, 
because the reddening correction (with the standard extinction 
coefficient $R_V=3.1$) nearly cancels the change in the absolute 
magnitude due to the change in the effective temperature as 
computed from the dereddened color index. We note that using 
higher metallicity of $-1.33$ for the primary component results in 
an increase of $0.05$~mag in the distance modulus. 

When the above temperature scale is applied in computing the distance 
with the aid of RRd stars, we get $DM=13.76$. This is lower than the 
above quoted value, because the zero point of the $T_{\rm eff}$ scale 
of Sekiguchi \& Fukugita (2000) is lower by $\Delta \log T_{\rm eff}=0.0044$ 
than the one used in our studies. 

The difference of $0.30$~--~$0.38$~mag (depending on the assumed 
[Fe/H] and $E_{B-V}$ of OGLEGC17) between the RRd and binary distances 
is very large and indicates that some basic problems may exist in one or 
both of these methods. However, we would like to mention that we get 
almost a factor of two smaller differences between the distance moduli, 
if we use the surface brightness formula given by T01 and used in the 
derivation of their preferred distance. With $E_{B-V}=0.13$, T01 give 
a value of $(V-K)_0=1.400$ for the infrared color index of the primary 
component. With this, and with the above quoted radius value one gets 
an absolute visual magnitude of $3.35$~mag. This yields a true distance 
modulus of $13.71$. If we use a lower reddening of $0.09$, we end up 
basically with the same figure, i.e., with $13.69$. In order to apply 
the surface brightness method based on the $V-K$ color index to our 
data set of the RRd stars with $B-V$ index, we computed corresponding 
$V-K$ indices by requiring that the two indices yield the same $T_{\rm eff}$. 
(We note that this procedure is independent of the temperature zero 
point.) In this way we derived a distance modulus of $13.88$~mag. This 
is still larger that the one obtained from the binary method, but the 
difference is only $-0.19$~mag (if we compare the results obtained with 
the same reddening). We suspect that the larger difference between the 
the distances derived from the surface brightness and 
$(B-V)_0\rightarrow T_{\rm eff}$ calibration in the binary method indicates 
that there might be some error in the color indices derived for the 
components by T01.


\section{The metallicity distribution of the RRab stars in $\omega$~Cen}

Individual metallicity values of RRab stars in $\omega$~Cen were 
calculated by employing Eq.~(4) on the sample used by KW01. The variable 
V137 was omitted because of its somewhat discordant light curve. For the 
remaining 43 stars we have an average [Fe/H] of $-1.48$ with a standard 
deviation of $0.18$. (Please note that we use a metallicity scale tied  
to the one defined by the high dispersion spectroscopic measurements --- 
see Jurcsik 1995; Clementini et al. 1995.) The distribution of the [Fe/H] 
values are displayed in Fig.~3. For comparison, we also show the 
metallicity distribution of 39 RRab stars in M~5. We used the data set 
of KW01, after omitting 4 variables (V6, V29, V77, V963), which 
were left out of this test for various resons (e.g., strange shape of 
light curve, high noise, etc.).  
%
%
%
\begin{figure}
\plottwo{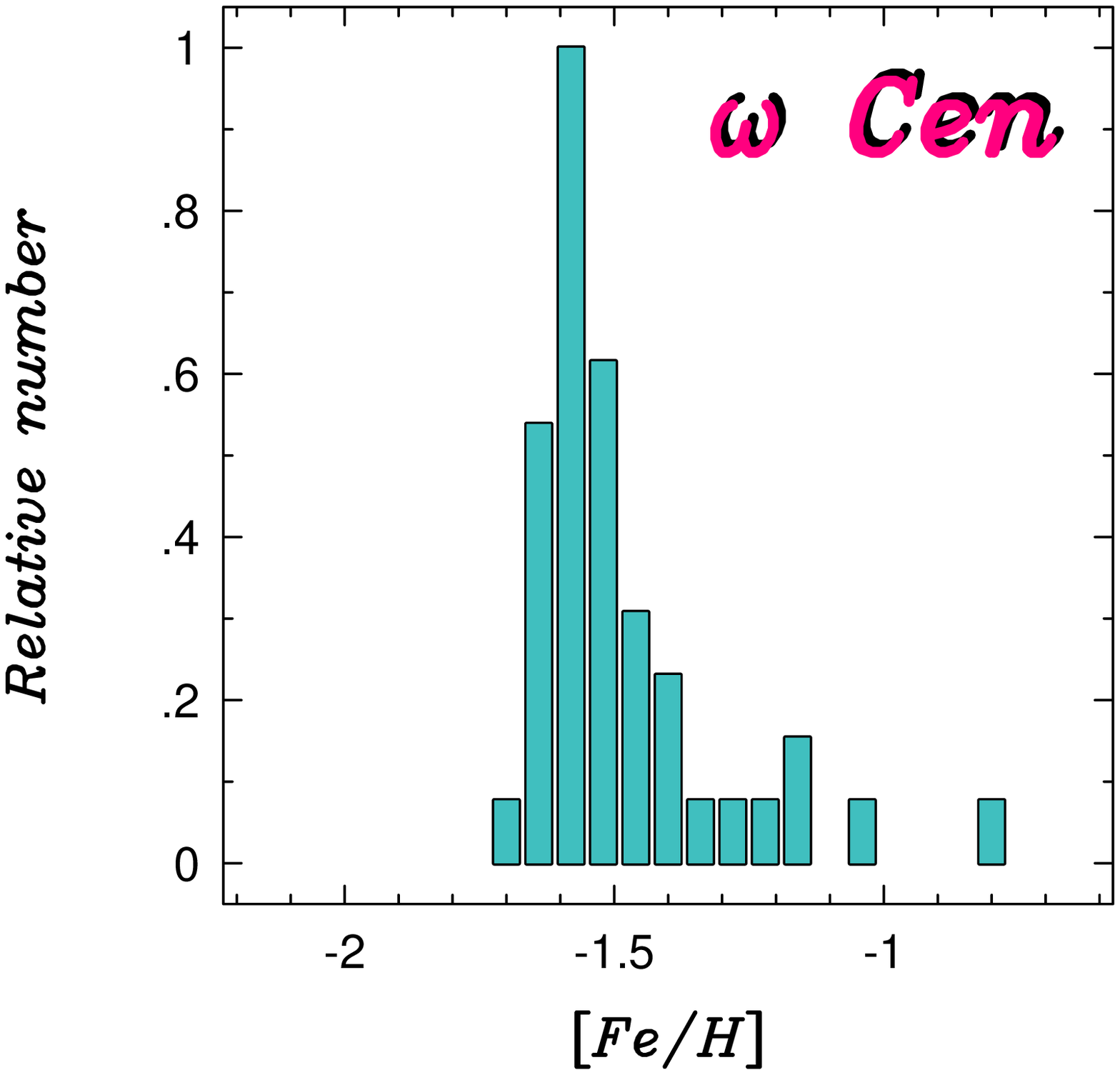}{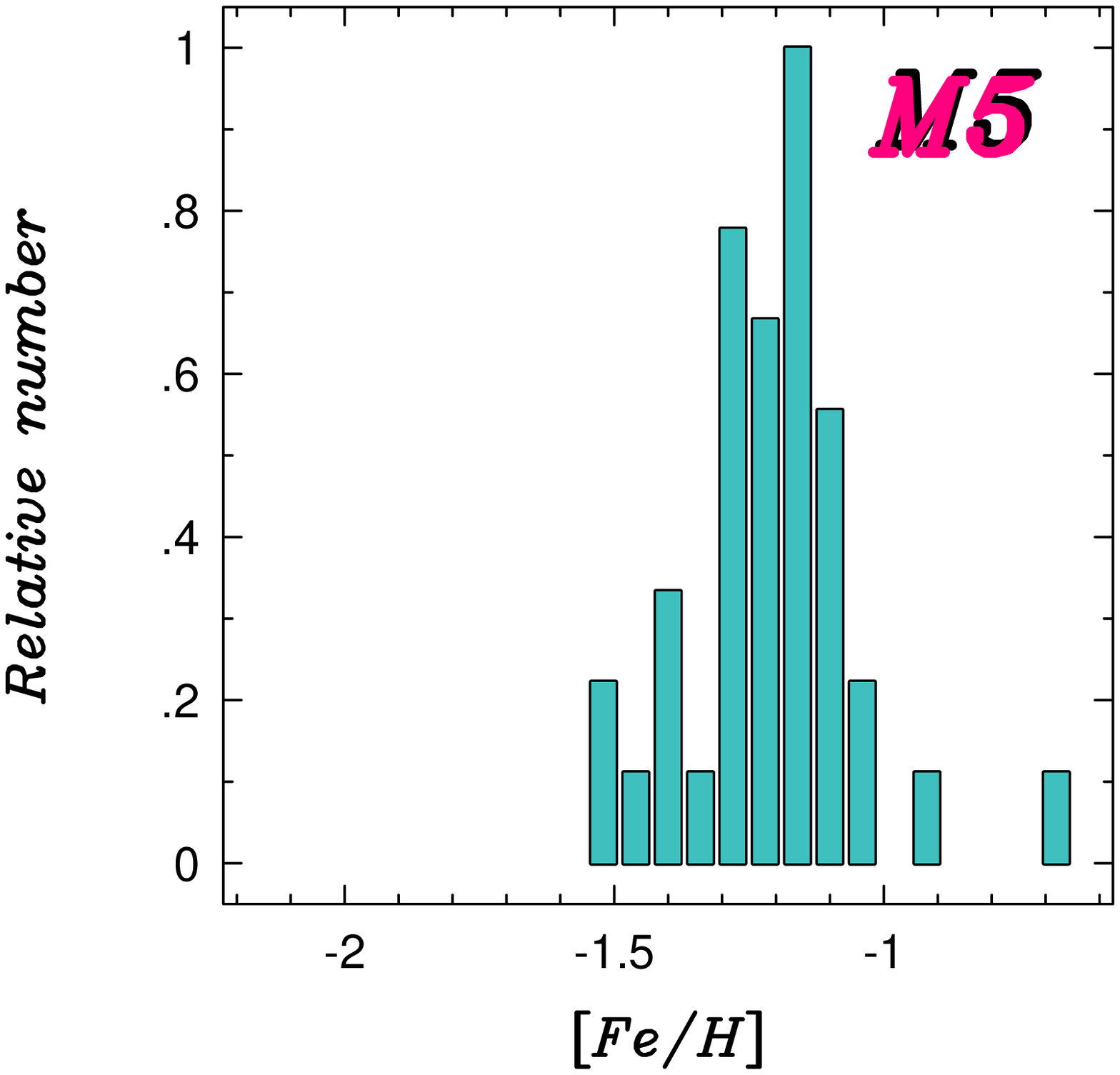}
\caption{Metallicity distributions of the RRab stars in two clusters}
\end{figure}
The average of [Fe/H] for M~5 is $-1.19$, with a standard deviation of 
$0.18$, which is the same as that of the $\omega$~Cen variables. This, 
and the distribution functions show that the two clusters are fairly 
similar from this respect, although for $\omega$~Cen the distribution 
seems to be more skewed, in a similar way as the distribution obtained 
from the spectroscopic observations of giant stars by Suntzeff \& Kraft 
(1996). On the other hand, our [Fe/H] values seem to be even more 
narrowly distributed than those of the giants. We note that direct 
$\Delta S$ measurements of Butler et al. (1978) yielded a very wide 
distribution. Although the metallicity dispersion obtained from 
multicolor photometry by Rey et al. (2000) is narrower, it is still 
broader than ours, and does not show the asymmetric pattern visible 
in the distribution of the giants. It is suspected that the direct 
estimates obtained from photometry is still suffering from a 
considerable amount of noise.   

The bulk of the distribution of [Fe/H] for M~5 is about $0.2$~dex 
wider than the one obtained from the direct spectroscopic data on 
giants by Ivans et al. (2001). It is not known what is the reason 
of this difference. Although care was taken to avoid variables 
with amlitude changes or large scatter, some part of the larger 
[Fe/H] range might originate from hidden Blazhko variables not 
revealed by the limited data available for most of the stars. Nevertheless, 
it is interesting to note that the RRab variables of the two clusters 
show the same amount of metallicity dispersion, although only 
$\omega$~Cen is considered in general as an example for a globular 
cluster with measurable metallicity spread. We refer to KW01 for 
metallicity estimates obtained from the light curves of the RRab stars 
in additional globular clusters.


\section{Conclusions}

In this paper we summarized the method of determination of the physical 
parameters of RR~Lyrae stars based on the Fourier analysis of the light 
curves of fundamental mode variables and on the pulsation study of 
double-mode pulsators. The method is applied to the variables of 
$\omega$~Cen. The following results were obtained:

\begin{itemize}
\item
The overall reddening determined from two independent data sets yielded 
basically the same result $E_{B-V}=0.09$~mag with a standard deviation 
of $0.03$~mag. Considering both samples together, the statistical error 
of the above mean is $0.005$~mag. The zero point ambiguity of this 
reddening value is estimated to be less than $\pm(0.01$--$0.02)$~mag. 
\item
By using the above two data sets, we determined the relative distance 
modulus of $\omega$~Cen from globular clusters with known absolute 
distance moduli computed from their double-mode variables. The derived 
distances were consistent within $0.02$~mag and yielded a mean true 
(dereddened) distance modulus of $13.83$~mag with a statistical error 
lower than $0.01$~mag. The largest systematic error comes from the 
color index~$\rightarrow T_{\rm eff}$ calibration, which (based on the 
differences between the currently available $T_{\rm eff}$ calibrations), 
does not exceed $0.1$~mag. 
\item
A comparison was made between the above distance modulus and the one 
derived by Thompson et al. (2001) from a detached binary of this cluster. 
By using the same temperature calibration, we found our distance to 
be larger by $0.3$~--~$0.4$~mag, depending on the assumed reddening 
and metallicity of the binary. By using a surface brightness method 
as a substitute for the temperature calibration, the difference decreased 
to $0.2$~mag, with both methods giving longer distances. 
\end{itemize}


\acknowledgments
This work has been supported by the following grants: 
{\sc otka t$-$026031}, {\sc t$-$030954} and {\sc t$-$038437}.


\end{document}